# Resource Efficient LDPC Decoders for Multimedia Communication

Vikram Arkalgud Chandrasetty and Syed Mahfuzul Aziz

*Abstract*—**Achieving high image quality is an important aspect in an increasing number of wireless multimedia applications. These applications require resource efficient error correction hardware to detect and correct errors introduced by the communication channel. This paper presents an innovative flexible architecture for error correction using Low-Density Parity-Check (LDPC) codes. The proposed partially-parallel decoder architecture utilizes a novel code construction technique based on multi-level Hierarchical Quasi-Cyclic (HQC) matrix with innovative layering of random sub-matrices. Simulation of a high-level MATLAB model shows that the proposed HQC matrices have bit error rate (BER) performance close to that of unstructured random matrices. The proposed decoder has been implemented on FPGA. It is very resource efficient and provides very high throughput compared to other decoders reported to date. Performance evaluation of the decoder has been carried out by transmitting JPEG images over an AWGN channel and comparing the quality of the reconstructed images with those from other decoders.**

## I. INTRODUCTION

Multimedia communication is an integral part of rapidly increasing number of applications including iPads, mobile phones and other handheld devices. Consequently, there is a strong interest in creating high performance hardware architectures with small overhead to enable error correction in multimedia communication. This paper presents a resource efficient architecture for error correction using Low-Density Parity-Check (LDPC) codes.

LDPC codes have emerged as one of the most popular forward error correcting (FEC) technique that can achieve bit error rate (BER) performance close to Shannon Limit [1]. The inherent structure of the LDPC matrix provides high degree of parallelism and flexibility for designing a decoder for various applications – Worldwide Interoperability for Microwave Access (WiMax), Wireless Local Area Network (WLAN) and Digital Video Broadcasting - Satellite - Second Generation

(DVB-S2) [2]. A fully-parallel architecture implementation of an LDPC decoder provides very high throughput but requires large hardware resources to achieve this performance [3-5]. Also, the complexity of the decoder increases drastically with longer code lengths. Therefore, an alternate solution to this problem is to use resource efficient partially-parallel architecture [6]. This architecture uses only a few number of decoding nodes and reuses them iteratively in the process. Unlike that in a fully-parallel decoder, it also utilizes block memories (in an FPGA) to store and access intermediate extrinsic messages. However, the advantages of partially-parallel architecture are achieved by sacrificing the throughput of the decoder due to additional clock cycles required for processing [7].

A partially-parallel decoding architecture provides a trade-off between hardware requirements and throughput. The number of parallel nodes (check node and variable node) required by the decoder is based on the partition size of the matrix (also known as the base matrix). Also, the complexity of the addressing scheme required for handling intermediate messages substantially depends on the structure of the LDPC matrix. Therefore, the hardware requirement of a partially-parallel architecture based decoder predominantly relies on the structure and complexity of the LDPC matrix [8]. In order to alleviate the complexity of the decoder, structured Quasi-Cyclic (QC) [9] based matrix construction methods are widely used. This technique constructs an LDPC matrix by using an array of cyclically-shifted base matrices [10]. The parallelism factor of partially-parallel decoder architecture is normally defined by the size of the base matrix. Hierarchical QC (HQC) [11] matrices are constructed with several levels of sub-matrices, with the last level corresponding to the base matrix. HQC based technique has the flexibility for constructing LDPC matrices of variable code lengths and code rates [12]. However, not all QC based matrix leads to comparable decoding performance (BER and average iterations) to that of unstructured matrices [10]. Therefore, constructing an LDPC matrix that reduces the complexity of partially-parallel decoder and also achieve optimum decoding performance is a challenge.

In a wireless communication system, protection and reliable transmission of multimedia content is of paramount interest [13]. LDPC codes are used to protect uncompressed grayscale images from errors [14], [15]. For better protection of baseline Joint Photographic Experts Group (JPEG) images, an unequal error protection (UEP) [16] scheme using LDPC

Vikram Arkalgud Chandrasetty is with the School of Electrical and Information Engineering, University of South Australia, Mawson Lakes, SA 5095, Australia (phone: +61 8 8302 3241; fax: +61 8 8302 3384; e-mail: vikramac@ieee.org).

Syed Mahfuzul Aziz is with the School of Electrical and Information Engineering, University of South Australia, Mawson Lakes, SA 5095, Australia (e-mail: mahfuz.aziz@unisa.edu.au).

codes and Reed-Solomon (RS) codes are presented in [17]. Performance evaluation of hybrid combination of RS and LDPC codes in [18], [19] shows increased reliability in transmission of multimedia content.

This paper presents a 3-Level HQC (3L-HQC) matrix construction technique with Layered Permutation (LP) [20]. The 3-Levels of hierarchy in the matrix provide flexibility of generating LDPC codes of different code lengths and code rates. The matrix can also be easily configured for applications such as WiMax, WLAN and DVB-S2. The proposed matrix consists of a permuted matrix in the level-2 of the hierarchical structure. Different combinations of permuted random matrices are inserted in layers of the LDPC matrix to provide randomness in the matrix structure. Simulation results show that the proposed matrix has a marginal degradation in BER performance compared to the unstructured random matrices. It also outperforms the 2-Level HQC based LDPC decoders [11]. The HQC-LP technique with a brief discussion on high level FPGA architecture of the decoder is presented in [21]. A detailed presentation of the hardware architecture of the decoder and its operation is presented in this paper. In addition, it also presents the performance analysis of the proposed decoder in multimedia communication, particularly for images.

FPGA implementation of the partially-parallel architecture using the proposed 3L-HQC matrix with LP leads to significant reduction in memory requirements compared to other partially-parallel decoder architectures reported to date. In addition to that presented in [4], performance of the decoder has also been evaluated in this paper through simulations by transmitting and reconstructing JPEG images over an Additive White Gaussian Noise (AWGN) channel. A visual comparison of the reconstructed images against the original images shows that the quality of the reconstructed images is better at low BERs. The image quality improves when the proposed LDPC decoder is designed with longer code lengths.

The rest of the paper is organized as follows. A brief overview of unstructured and structured LDPC matrices along with decoder implementation complexity is presented in section 2. In section 3, the construction and performance analysis of the proposed matrix is presented. Applicability of the proposed matrix for various applications is also presented. It is then followed by a partially-parallel decoder implementation in section 4. Section 5 presents performance evaluation of the proposed decoder for multimedia communication.

## II. PROPOSED HQC MATRIX WITH LAYERED PERMUTATION

QC based techniques [9] are less flexible for constructing matrices of variable sizes, when compared to unstructured matrices. This limitation is due to the use of array of sub-matrices that are fixed in size. The technique proposed here is flexible for constructing matrices by exploiting the advantages of using HQC methods [11]. As opposed to the 2-Level hierarchy in HQC [12] the proposed technique introduces 3-Level hierarchy to efficiently organize the structure and construct flexible matrices with variable code lengths/rates. Also, Permuted sub-matrices are inserted in layers of the LDPC matrix. This introduces virtual randomness in the matrix, similar to that of unstructured matrices, to improve the decoding performance. The following sub-sections present a detailed explanation on the construction and analysis of the proposed technique.

### A. Construction of the Matrix

In order to illustrate the matrix construction process, a ½ rate (3, 6) regular LDPC matrix is considered in this example. A simple structure of the proposed 3L-HQC matrix with LP is shown in Fig. 1.

**Level-1:** The proposed matrix has 3-Levels of hierarchy. The first level of matrix in the hierarchy is termed as the Core matrix. This level is responsible for maintaining the rate and regularity of the LDPC matrix. For example, in case of ½ rate (3, 6) regular LDPC code configuration, the Core matrix (H) consists of 3 rows and 6 columns (see Fig. 1). Further down the matrix construction process, each of the elements in the Core matrix that are expanded maintains a regularity of (1, 1). This retains the overall regularity of (3, 6) in the LDPC matrix.

**Level-2:** The second level of the matrix is obtained by expanding each of the elements in the Core matrix with a circularly shifted identity matrix (L) of size 'N', similar to [22]. However, this matrix (L) is again expanded by placing an array of circularly shifted Permuted matrices (Rx) of size 'R'. A Permuted matrix is constructed by placing a positive integer value randomly in the matrix. Examples of integer values are shown as subscript of 'I' in Fig. 1. This level of the matrix structure is predominantly responsible for expansion and construction of LDPC matrices with various code lengths for a particular application.

Note that different combinations of Permuted matrices are used in layers (each rows of Core matrix) of the LDPC matrix. The subscripts in each of the elements in the Core matrix (H) illustrate the layering of the Permuted matrix. For example, a subscript of (x, y) indicate that an 'xth' combination of Permuted matrix is used for expansion of that particular element in the Core matrix with a circular shift of 'y'.

**Level-3:** In the third level, each of the non-zero elements in the Permuted matrix is expanded by a Base matrix (I). This matrix is a circularly shifted identity matrix of size 'P'. The number of circular shifts in a Base matrix depends on the elements in the Permuted matrix. This is indicated by the subscript of 'I' in the Permuted matrix, as shown in Fig. 1. The size of the Base matrix defines the parallelism factor (P) of the LDPC decoder. That is, the number of check nodes and variable nodes required for parallel processing.

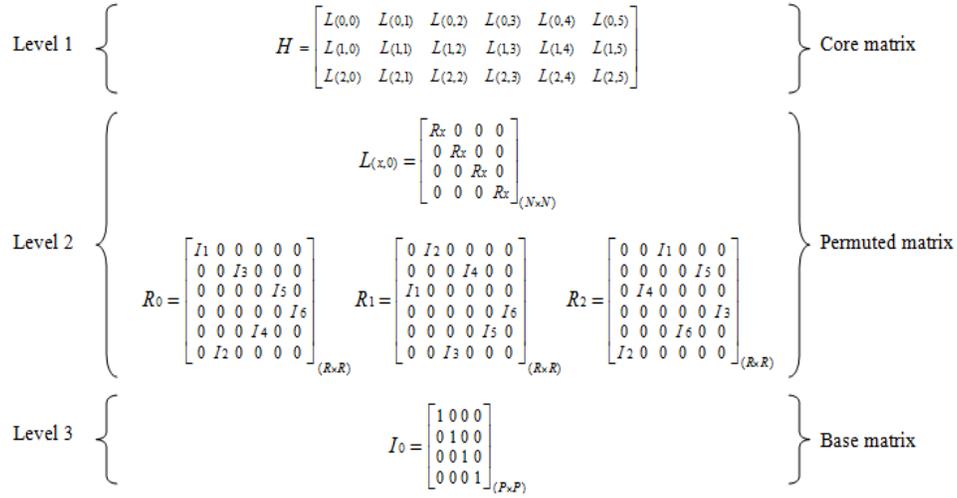

Fig. 1. Illustration for constructing the proposed 3L-HQC matrix with LP

## B. Various Matrix Configurations

The proposed technique can be configured to generate LDPC matrices with different code lengths by varying the 'N', 'R' and 'P' parameters. Some of the possible configurations that are suitable for WiMax [23], WLAN [24] and DVB-S2 [25] applications are shown in Table I.

Note that a number of decoders have been proposed that uses a flexible multi-rate and multi-length LDPC matrix [12, 26, 27]. However, the proposed matrix is more flexible for constructing LDPC matrices for multiple applications (as shown in Table I) without compromising the decoding performance. This flexibility is possible due to the additional level (3rd) in the LDPC matrix hierarchical structure.

TABLE I
CONFIGURATIONS OF THE PROPOSED MATRIX FOR VARIOUS APPLICATIONS

| WiMax (P=16) | | | | WLAN (P=18) | | | | DVB-S2 (P=27) | | | |
|---|---|---|---|---|---|---|---|---|---|---|---|
| CL | CR | R | N | CL | CR | R | N | CL | CR | R | N |
| 576 | 1/2 | 6 | 1 | 648 | 1/2 | 6 | 1 | 16200 | 1/3 | 5 | 20 |
| 672 | 1/2 | 7 | 1 | 1296 | 1/2 | 6 | 2 | 16200 | 2/3 | 5 | 20 |
| 768 | 1/2 | 8 | 1 | 1944 | 1/2 | 6 | 3 | 64800 | 1/2 | 8 | 50 |
| 864 | 1/2 | 9 | 1 | 648 | 2/3 | 6 | 1 | 64800 | 1/3 | 8 | 50 |
| 960 | 1/2 | 10 | 1 | 1296 | 2/3 | 6 | 2 | 64800 | 2/3 | 8 | 50 |
| 1056 | 1/2 | 11 | 1 | 1944 | 2/3 | 6 | 3 | 64800 | 5/6 | 8 | 50 |
| 1152 | 1/2 | 6 | 2 | 648 | 5/6 | 6 | 1 | - | - | - | - |
| 1728 | 1/2 | 6 | 3 | 1296 | 5/6 | 6 | 2 | - | - | - | - |
| 2304 | 1/2 | 6 | 4 | 1944 | 5/6 | 6 | 3 | - | - | - | - |

Note: CL = Code Length; CR = Code Rate;

## C. Performance Analysis using a High-Level Model

To analyze the decoding performance of the proposed matrix (3L-HQC with LP), simulations were carried out and compared against 2L-HQC and PEG based matrices. A software simulation model was developed using C programs and executed in the MATLAB environment [28]. A ½ rate (3,

6) regular 2304-bit LDPC code (WiMax) was used to assess the BER and average iterations for different matrices. A Modified Min-Sum (MMS) algorithm [5] was used to reduce the hardware complexity and memory requirements [29]. For the simulations, the encoded data is assumed to have Binary Phase Shift Keying (BPSK) modulated and passed over an Additive White Gaussian Noise (AWGN) channel. The maximum number of iterations for the algorithm was set to 10.

The BER performance and average iterations against Eb/No (Signal strength per bit to Noise ratio) obtained from simulations are shown in Fig. 2 and Fig. 3 respectively. From Fig. 2, it is clear that the proposed matrix outperforms the 2L-HQC by 0.4 dB at a BER of $10^{-6}$. The PEG based random matrix has a marginal performance gain of less than 0.1 dB over the proposed matrix at a BER of $10^{-6}$. In case of average iterations (Fig. 3), the proposed matrix requires fewer iterations compared to 2L-HQC, while requiring more iterations compared to the PEG based matrix.

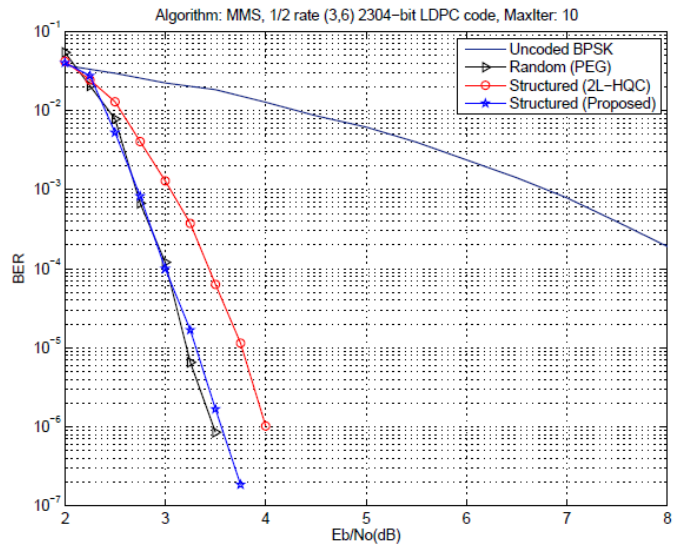

Fig. 2. Software simulation of BER performance for various matrices

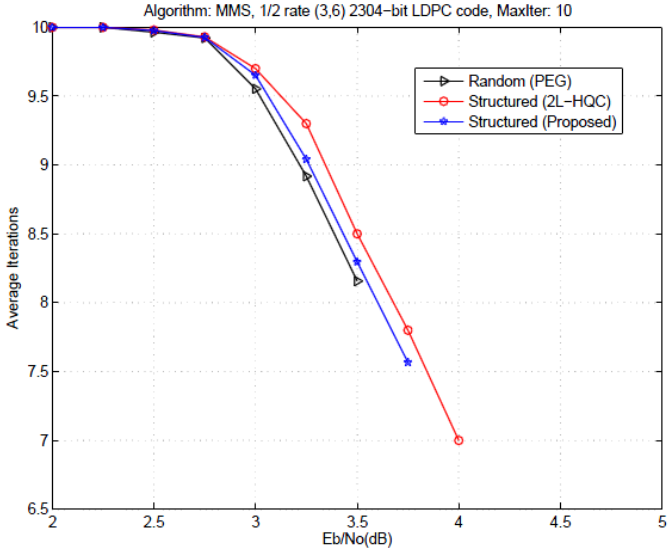

Fig. 3. Software simulation of average iterations for various matrices

## III. Hardware Modeling and Analysis

### A. Hardware Design of the Decoder

A prototype hardware model of an LDPC decoder using the proposed matrix has been designed using Verilog Hardware Description Language (HDL). In order to verify the feasibility of hardware implementation of the proposed matrix, the decoder model uses simple and straight-forward partially-parallel architecture.

As with the high-level simulation model in Section II, the hardware model of the decoder is designed for a ½ rate (3, 6) regular LDPC code, with code lengths 576, 1152 and 2304, which are compliant with WiMax applications (see Table I). The same Modified Min-Sum (MMS) algorithm [5] is used, which simplifies check node operation and uses reduced extrinsic message quantization [29].

A top-level block diagram of the hardware model of the LDPC decoder is shown in Fig. 4. The decoder consists of two major blocks: Decode Controller (DC) and Decode Processor (DP). The DC is responsible for controlling the decoding process and responding to external control signals. It also organizes and sequences the input data for decoding and to output the decoded data. The DP is responsible of the decoding process. It consists of Variable Node Processing Unit (VNPU), Check Node Processing Unit (CNPU), Variable Nodes (VN), Check Nodes (CN), Intrinsic Message Block (IMB) and the Permuted Matrix Memory Block (PMMB). Based on number of parallel nodes (P) for this configuration of the decoder, the VN and CN blocks consist of chain of 96 variable nodes and check nodes respectively (see Table I). The Permuted matrix information is stored in the form of Look-Up Tables (LUT) in PMMB. The VNPU and CNPU use these LUTs for accessing and storing messages at appropriate locations in the Block RAMs (BRAM). To start with the decoding process, the VNPU first accesses the intermediate message decoding data (extrinsic messages) from the BRAM and passes on to the VN. The VN processes this data along with the intrinsic message from IMB. The updated message is then passed to CNPU to be stored back in the BRAM. This cycle continues till all the variable nodes are processed for the entire code length of the decoder. Next, a similar message updating process is performed by CNPU and CN. This processing cycle of VNPU and CNPU completes a single decoding iteration of the decoder. The decoding process is stopped by DC when the maximum iteration is reached or the parity check is satisfied.

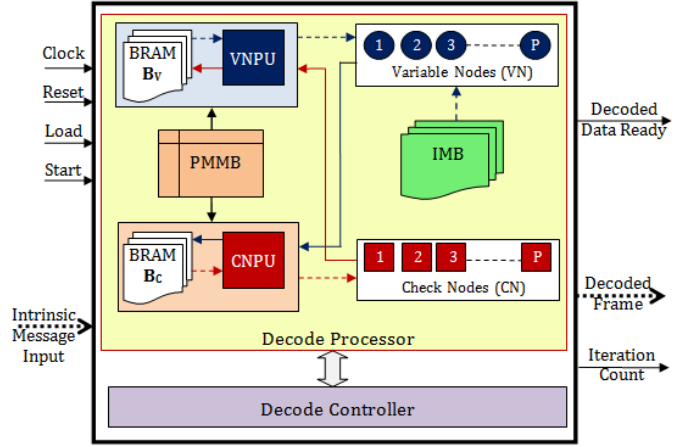

Fig. 4. Top level block diagram of the prototyped LDPC decoder

To start with the decoding process, the VNPU first accesses the extrinsic messages from the Block RAM ($B_V$) and passes them on to the VN. The VN processes this data along with the intrinsic messages from IMB. The extrinsic messages generated by the variable nodes are sent to CNPU in a pipelined fashion for updating them in the Block RAM ($B_C$). The timing diagram in Fig. 5 illustrates the sequence of operations performed when the variable node processing (VNP) cycle is active. Each of the VNPU message processing cycles (VN$_J$) indicates $P$ variable nodes operating in parallel. CM$_J$ indicates that $P$ number of extrinsic messages from the variable nodes is updated in the BRAM ($B_C$). The VNP is active for 'J' clock cycles until all the variable nodes are processed for the entire code length. The number of clock cycles 'J' for the complete VNP operation is given by (1).

$$J = \frac{CodeLength}{ParallelNodes(P)} \qquad (1)$$

When VNP operation is complete, a similar message updating process is performed by CNPU. The CNPU accesses the extrinsic messages from the BRAM ($B_C$) and passes them on to the CN. The extrinsic messages generated by the check nodes are pipelined and sent to the VNPU for updating it in BRAM ($B_V$). The CN also outputs parity check information to the DC. The timing diagram in Fig. 6 illustrates the sequence of operations performed when the CNP cycle is active.

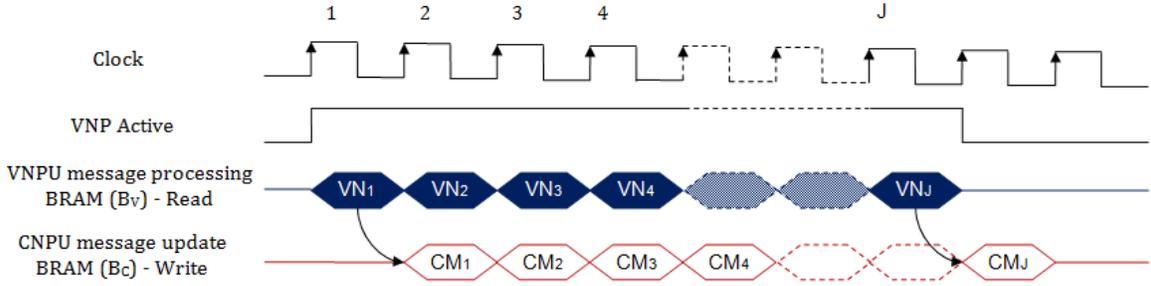

Fig. 5. Timing diagram illustrating variable node processing unit operation

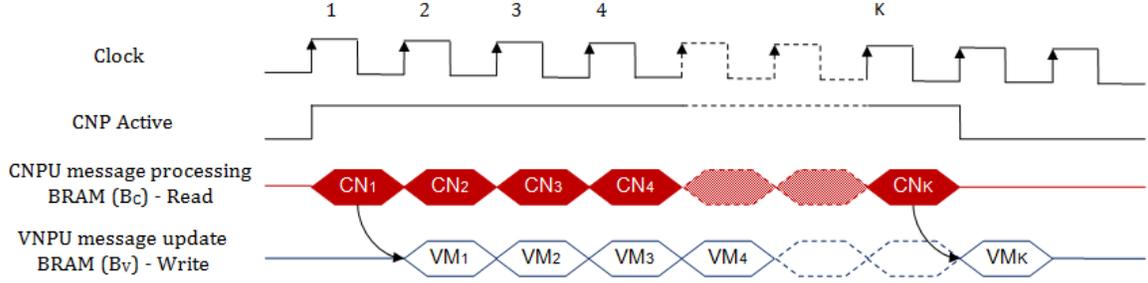

Fig. 6. Timing diagram illustrating check node processing unit operation

Each of the CNPU message processing cycle $CN_K$ indicates $P$ check nodes operating in parallel. $VM_K$ indicates that $P$ number of extrinsic messages from the check nodes is updated in BRAM ($B_V$). The CNP is active for 'K' clock cycles until all the check nodes are processed. The number of clock cycles 'K' for the complete CNP operation is given by (2).

$$K = \frac{CodeRate \times CodeLength}{ParallelNodes(P)} \qquad (2)$$

The combined processing cycles of VNPU and CNPU constitute a single decoding iteration of the decoder. The decoding process is stopped by the Decode Controller (DC) when the maximum iteration count is reached or the parity check is satisfied. The decoder requires additional clock cycles to compensate for the delays in VN and CN operations due to pipelined processing in each of the decoding iterations. The *latency (L)* of the decoder is '6' clock cycles and is constant for any code lengths or parallelism factor. The total number of clocks per decoding iterations ($N_{it}$) for the proposed decoder is computed by (3).

$$N_{it} = J + K + L \qquad (3)$$

For example, for a decoder using ½ rate 2304-bit LDPC code with 96 parallel nodes, $N_{it}$ is computed as follows:

$$J = \left(\frac{2304}{96}\right) = 24$$

$$K = \left(\frac{1/2 \times 2304}{96}\right) = 12$$

$$N_{it} = 24 + 12 + 6 = 42 \qquad (4)$$

### B. Analysis of Implementation Results

The hardware model of the proposed decoder has been simulated to determine performance. Fig. 7 and 8 show the BER performance and average iterations of the decoder respectively. As expected, the BER performance improves as the code length of the decoder increases (Fig. 7) at the cost of increased average iterations (Fig. 8). The hardware model of the decoder has been synthesized, placed and routed for implementation on a Xilinx Virtex 4 FPGA (XC4VLX160). The results obtained from synthesis and implementations have been used to summarize the decoder's hardware requirements and performance in Table II.

Table II also summarizes the hardware requirements and performance of other partially-parallel decoders reported in the literature. Among the partially-parallel decoder architectures reviewed [30-38], only those with configuration similar to the proposed decoder are listed in Table II. For example, each decoder in Table II has 96 nodes in parallel, is designed for a code length of 2304 and is compliant with the WiMAX (IEEE 802.16e) standard.

TABLE II
COMPARISON OF HARDWARE REQUIREMENTS AND PERFORMANCE OF VARIOUS DECODERS

| | Proposed | [30] | [31] | [32] | [33] | [34] |
|---|---|---|---|---|---|---|
| Application | WiMAX – IEEE 802.16e standard | | | | | |
| Code structure | 3L-HQC with LP | 2L-HQC | Irregular | NA | NA | PEG-QC |
| Parallel nodes (P) | 96 | | | | | |
| LUTs | 31,305 | 33,226 | 11,028 | 19,000 | 27,850 | 17,259 |
| Registers | 4,066 | 32,619 | 6,330 | 10,000 | 9,806 | 6,598 |
| BRAMs | 160 | 75 | 100 | 92 | NA | NA |
| Total memory (bits) | 20,736 | NA | 60,288 | NA | 100,552 | 271,104 |
| Clock frequency (MHz) | 82 | 192.4 | 110 | 160 | 100 | 155 |
| Avg. Throughput (Mbps) | 300 | NA | 278 | 10.4 | 154 | 232.5 |
| FPGA device | Virtex 4 | Virtex 4 | Virtex 2 | Virtex 5 | Stratix 2 | Stratix 2 |

NA: Data not available

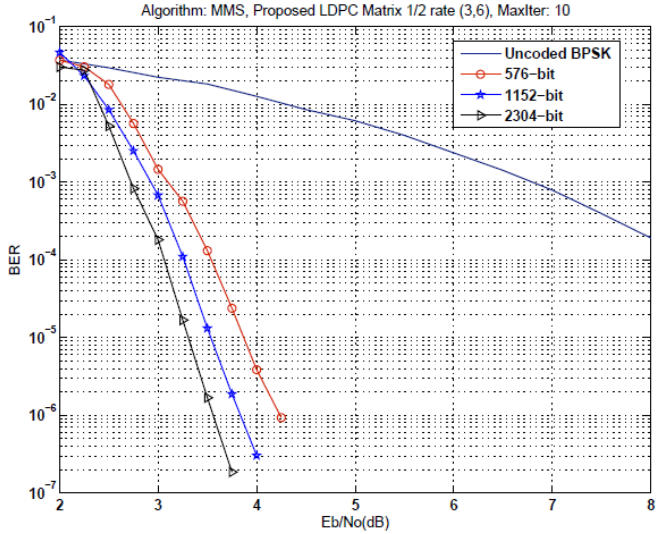

Fig. 7. BER performance of the proposed LDPC decoder from FPGA

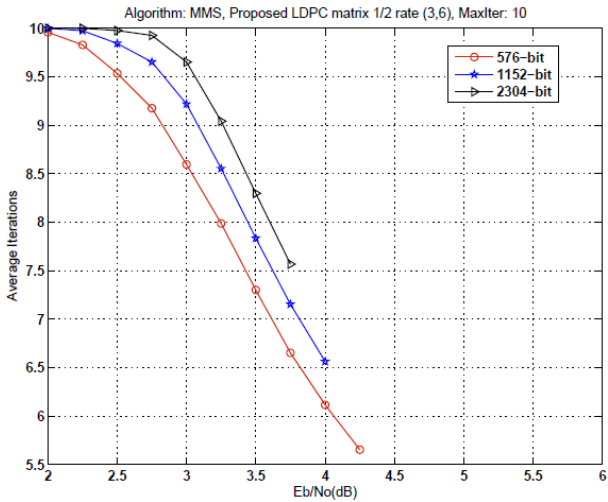

Fig. 8. Average iterations for the proposed LDPC decoder from FPGA

The *throughput (T)* of the implemented 3L-HQC decoder is computed using the formula given in (5). For a code length of 2304, the number of parallel nodes in the implemented decoder is 96. From (4), the total number of clock cycles per decoding iteration ($N_{it}$) is 42. At a maximum operating frequency of 82 MHz (obtained from the implementation results), the average *throughput* of the decoder using average iterations of 7.5 (see Fig. 8 at 3.75 dB $E_b/N_o$) is approximately 300 Mbps.

$$T = \frac{CodeRate \times CodeLength \times MaxOper.Frequency}{DecodingIterations \times N_{it}} \quad (5)$$

It is clear from Table II that the *throughput* of the proposed 3L-HQC decoder is significantly greater than all other decoders. This is achieved by incorporating an efficient pipelined and parallel processing scheme at the nodes (described in Section III A). It requires much less registers and significantly lower number of memory storage bits. The number of LUTs required is also less when compared to the 2L-HQC decoder [30]. Although the decoders presented in [31] and [32] use less LUTs, their throughputs are significantly lower and they require much higher number of registers and memory bits.

Comparing the hardware requirements of the proposed decoder implemented on the Xilinx Virtex 4 FPGA against those implemented on the Altera Stratix II FPGA [33, 34] is not very straightforward, because these two devices have different structures. However, comparison of the LUTs, registers and memory bits required will provide a reasonable indication of the hardware requirements on either device. Although the proposed 3L-HQC decoder uses larger number of LUTs compared to [33, 34], it requires less number of registers and significantly less memory storage bits. In addition, the proposed decoder has significantly greater throughput.

The throughput of the proposed decoder is easily scalable by increasing the parallelism factor ($P_f$). However, this results in an increase in the hardware and memory requirements. A comparison of the hardware requirements and throughput of the proposed 3L-HQC decoder for different values of $P_f$ is shown in Table III. The presented data is for a ½ rate (3, 6) regular 2304-bit LDPC decoder implemented on a Xilinx Virtex 5 FPGA. The number of parallel check nodes and variable nodes are equal to P. Note that the memory requirement (in bits) for the decoder is constant for a given code length even though the parallelism factor is changed.

TABLE III
HARDWARE REQUIREMENTS AND PERFORMANCE OF THE PROPOSED 3L-HQC
DECODER FOR VARIOUS PARALLELISM FACTORS

| Parallel factor ($P_f$) | 1 | 2 | 3 | 4 |
|---|---|---|---|---|
| Parallel nodes (P) | 16 | 48 | 96 | 144 |
| Slices | 1137 | 3141 | 5583 | 8430 |
| LUTs | 3522 | 9547 | 18542 | 27558 |
| Registers | 847 | 2024 | 3992 | 5961 |
| BRAMs (18K) | 29 | 87 | 160 | 232 |
| Memory (bits) | 20736 | | | |
| Clock (MHz) | 162 | 144 | 126 | 114 |
| Clocks per decoding iteration | 222 | 78 | 42 | 30 |
| Average Throughput (Mbps) | 104 | 266 | 432 | 548 |
| FPGA device | Xilinx Virtex 5 (XCVLX110T) | | | |

## IV. PERFORMANCE EVALUATION FOR MULTIMEDIA COMMUNICATION

### A. Evaluation Technique

Performance evaluation has been carried out by transmitting images over an AWGN channel and reconstructing the images using the proposed LDPC decoder in MATLAB environment. The evaluation scheme is illustrated in Fig. 9. Although JPEG has standardized error correcting features [39], the received images may still be distorted due to errors during transmission. Therefore, a hybrid encoding technique presented in [19], [40] is incorporated in the communication system to ensure reliable transmission of image data. This technique uses Reed-Solomon (RS) codes for encoding header and tail sections of the JPEG image [41]. The RS encoded headers along with the rest of the JPEG data is then encoded using LDPC codes [18], [42]. The RS-LDPC encoded data is transmitted over an AWGN channel, where the data is deliberately subjected to some errors. The received erroneous data is first decoded using the proposed LDPC decoder. Then the header/tail sections of the JPEG image are decoded using a RS decoder for image reconstruction. The loss in compression rate of the image due to introduction of RS coding scheme is negligible when compared to the data integrity of multimedia content achieved using such encoding technique [17].

### B. Comparison of Performance

The quality of the reconstructed JPEG image is compared against the original transmitted image under various BER conditions and different code lengths of the proposed LDPC decoder [43]. Colored image samples of size 512×512 pixels compressed using JPEG2000 [44] standard were used for simulations. For a decoder with a code length of 2304, visual comparison of the quality of original and reconstructed images for different decoders is shown in Table IV. The images reconstructed using PEG based and proposed decoder have negligible difference compared to the original images. However, for 2L-HQC based decoder there is a slight drift in the luminance component of the images.

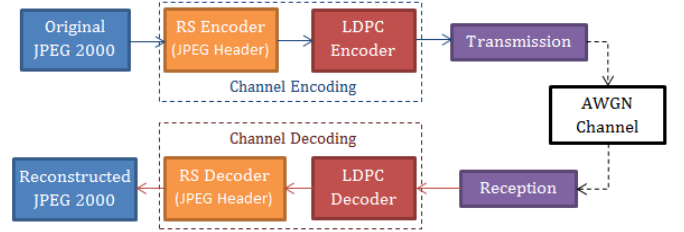

Fig. 9. Block diagram of performance evaluation technique

The quality of reconstructed images at various BER conditions is also analyzed by computing peak signal to noise ratio (PSNR) with respect to the original image [14], [45]. The PSNR is calculated using the following formula (6) and (7):

$$PSNR(dB) = 10 \times \log\left(\frac{P_{max}^2}{MSE}\right) \qquad (6)$$

$$MSE = \sum_{i=1}^{x}\sum_{j=1}^{y}\frac{\left(\left|A_{ij}-B_{ij}\right|\right)^2}{x \times y} \qquad (7)$$

where, MSE: Mean-Square Error
  $P_{max}$: Maximum value of a pixel in the image
  A: Pixel value of original image
  B: Pixel value of reconstructed image
  x: Height of the image in pixels
  y: Width of the image in pixels

The BER versus PSNR plots for various LDPC matrices using a JPEG image sample (Lena) are shown in Fig. 10. The PSNR values for PEG and the proposed matrix are similar over the BER range of $10^{-4}$ to $10^{-6}$. However, 2L-HQC based LDPC matrix has comparatively lower PSNR values over the same BER range. The same aspect has been verified by analyzing the visual quality of the reconstructed images presented in Table IV.

The BER versus PSNR plots for three different reconstructed image samples are shown in Fig. 11. Clearly, the PSNR values are higher for images transmitted at lower bit error conditions. For example, PSNR > 70 dB at a BER of $10^{-6}$ and PSNR < 50 dB at a BER of $10^{-4}$.

The quality of the reconstructed images has also been evaluated for different code lengths of the proposed LDPC decoder. The code length versus PSNR plots are shown in Fig. 12. It is clear from this figure that using decoders with larger code lengths achieves higher PSNR values and hence better quality of the reconstructed images.

TABLE IV
COMPARISON OF THE QUALITY OF ORIGINAL AND RECONSTRUCTED IMAGES

| Original image | Reconstructed image at $E_b/N_o$ of 3.5 dB using 2304-bit LDPC code for | | |
|---|---|---|---|
| | Proposed decoder | PEG based decoder | 2L-HQC based decoder |
| (a) Lena | | | |
| (b) Baboon | | | |
| (c) GoldenGate | | | |

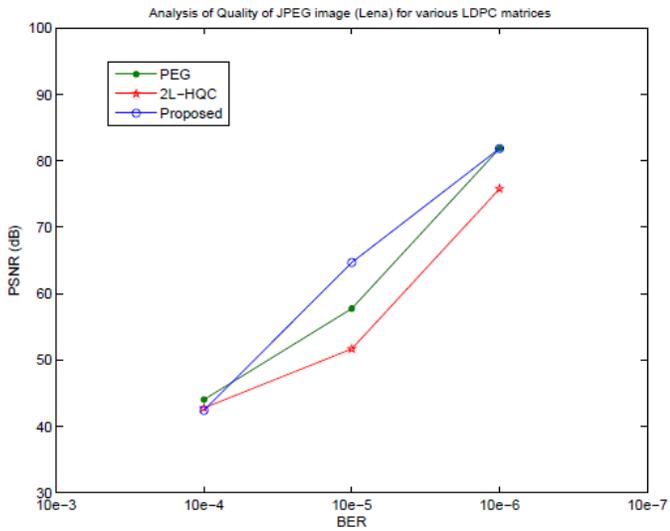

Fig. 10. BER versus PSNR for Lena using various LDPC matrices

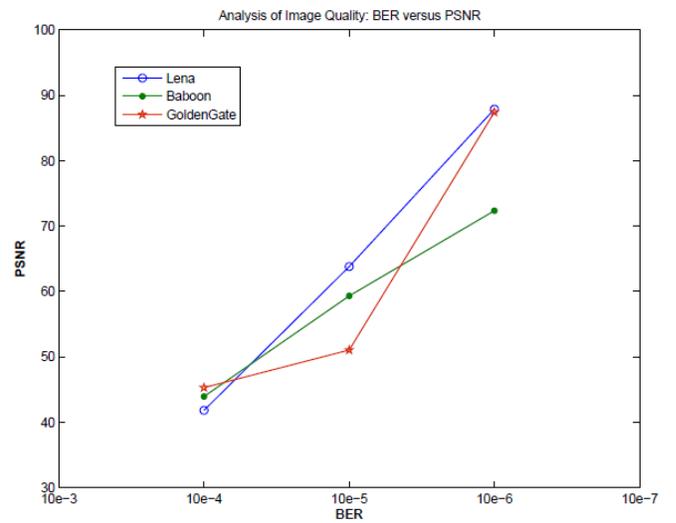

Fig. 11. BER versus PSNR for different reconstructed image samples

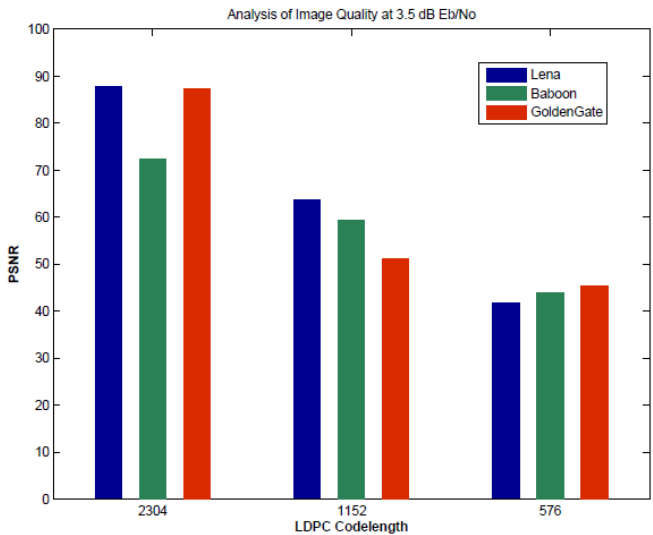

Fig. 12. LDPC code length versus PSNR for the reconstructed images

## V. CONCLUSION

This paper has presented a resource efficient decoder suitable for error correction in applications involving multimedia (image) communication. It relies on a novel technique to flexibly construct LDPC matrices for different code lengths using a Hierarchical Quasi-Cyclic (HQC) based approach. It is shown that using multi-level hierarchy and innovative layered permutation leads to (1) flexibility in code construction, (2) BER performance close to Progressive Edge Growth (PEG) based matrices, (3) reduced hardware implementation complexity, (4) better controllability of parallelism factor and (5) scalable throughput. A ½ rate (3, 6) regular 2304-bit LDPC decoder implemented using the proposed 3L-HQC matrix achieves a throughput of 300 Mbps, which is much higher than other reported decoders having the same specifications. It uses less LUTs than 2L-HQC, and significantly less registers and memory storage bits compared to all the reported decoders. The latter will easily offset the moderately higher LUT count of the proposed 3L-HQC decoder compared to some of the reported decoders. Simulations were carried out to assess the quality of JPEG images transmitted over an AWGN channel and reconstructed after error correction using various decoders. It is shown that the proposed matrix delivers same quality images as PEG based matrix and better quality images than 2L-HQC.

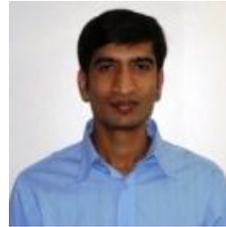

**Vikram Arkalgud Chandrasetty** received Bachelor Degree in Electronics and Communication Engineering from Bangalore University (India) in 2004 and Master Degree in VLSI System Design from Coventry University (UK) in 2008.

He was working with Core Networks Division at Motorola India as Software Engineer (2005-2007), where he was part of the billing and call processing R&D team of Motorola Soft-Switch (MSS) for Mobile Switching Centres (MSC). He also worked for SoftJin Technologies as Senior Software Engineer (2007-2008) focusing on Electronic Design Automation (EDA) and FPGA applications design. He was involved in the design and development of Programmable Synthesis Engine (PSE) for custom FPGA architectures and structured ASICs. He was also working on software modelling and FPGA implementation of Motion Estimation algorithms for H.264 Advance Video Coder.

Mr Vikram is currently working towards his doctoral thesis at the School of Electrical and Information Engineering, University of South Australia. He is exploring low complexity algorithms for decoding LDPC codes and investigating efficient architectures for hardware implementation. His research is mainly focused on implementing high performance LDPC decoders on reconfigurable devices.

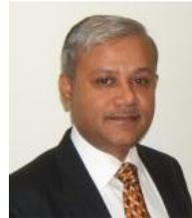

**Syed Mahfuzul Aziz** received Bachelor and Masters Degrees, both in electrical & electronic engineering, from Bangladesh University of Engineering & Technology (BUET) in 1984 and 1986 respectively. He received a Ph.D. degree in electronic engineering from the University of Kent (UK) in 1993 and a Graduate Certificate in higher education from Queensland University of Technology, Australia in 2002.

He was a Professor in BUET until 1999, and led the development of the teaching and research programs in integrated circuit (IC) design in Bangladesh. He joined the University of South Australia in 1999, where he is currently an associate professor. In 1996, he was a visiting scholar at the University of Texas at Austin when he spent time at Crystal Semiconductor Corporation designing advanced CMOS integrated circuits. He has been involved in numerous industry projects in Australia and overseas, and has attracted funding from reputed research organisations such as the Australian Research Council (ARC), Australian Defence Science and Technology Organisation (DSTO), and Cooperative Research Centre, Australia. He has authored over 110 refereed research papers. His research interests include digital CMOS IC design and testability, modelling and FPGA implementation of high performance processing systems, biomedical engineering and engineering education.

Prof Aziz is a senior member of IEEE and a member of Engineers Australia. He has received numerous professional and teaching awards including the Prime Minister's Award for Australian University Teacher of the Year (2009). He has served as member of the program committees of many international conferences. Among the journals he has reviewed in the last five years are the *IEEE Transactions on Computer, IEEE Transactions on Image Processing, IEEE Transactions on Education, IEEE Communications Letters, Electronics Letters, Computers & Electrical Engineering – An International Journal*.